\documentstyle[twocolumn,aps]{revtex}
\begin{document}
\title{ Large-Scale Synchrony in Weakly Interacting Automata}

\author{ Eric J. Friedman\dag  and
         A.S. Landsberg\ddag }
\address{
 {\dag}Rutgers University,
New Brunswick, NJ 08903\\
{\ddag}W.M. Keck Science Center,
The Claremont Colleges,
Claremont, CA 91711}

\maketitle
\begin{abstract}
We study the behavior of two spatially distributed (sandpile)
models which are weakly
linked with one another.  Using a Monte-Carlo implementation of
the renormalization group and algebraic methods, we
describe how large-scale correlations emerge between the two systems,
leading to synchronized behavior.

\end{abstract}

\section{Introduction}
Interacting systems have long been the subject of considerable interest and study,
and they can display a rich variety of behaviors.  One of the premier
concepts that has
emerged from such studies is the notion of synchronization, both in its
traditional sense
 as well as in one of its modern variants
(e.g., phase synchronization, lag synchronization, etc.)
\cite{Rosenblum96-97,Kocarev96,Rulkov95,Pecora90}. A particularly
interesting class of systems to consider in light of these broadening
notions of synchronization is provided by automata (``sandpile'') models.
Automata offer a rich assortment of well-studied, complex behaviors (e.g.,
self-organized criticality \cite{Bak87-88}), and have been used extensively
in the literature to model a variety of  physical phenomena (e.g.,
\cite{SOC}). If two such automata are permitted to weakly interact \cite{econnote}, an
interesting type of synchronization effect is seen to emerge:
While small events on one sandpile are essentially uncorrelated with small
events on the other, large-scale
events are so highly correlated that not only is a large event on one
 sandpile almost
always
concomitant with a large event on the other sandpile, but
the two events are in fact approximately
equal
in magnitude (with rms fractional deviation approaching zero). (See Fig.\ 1.)  This result holds despite the weakness of the coupling
between the sandpiles.  Note that this ``synchronization'' between  sandpiles is
not periodic (i.e., the time interval between synchronized large-scale events is not
fixed),
nor is it completely random (since correlations exist between temporal spacing of events
and event size.)

We can glean some basic insight into the origin of this form of
synchronization from a relatively simple plausibility argument: As a  large
avalanche sweeps across one sandpile, it will, owing to the weak coupling,
spill some small yet nontrivial number of grains onto the other sandpile.
Since sandpile models (like other self-organized critical systems) are
capable of generating avalanches of all sizes, these spilled grains could
conceivably induce a large subsidiary avalanche in the second sandpile. In
turn, this subsidiary avalanche could spill some grains back onto the first
sandpile, and so on.  In this manner, it is possible to imagine how high
levels of correlations might develop between the two sandpiles during large
events. It is thus reasonable to conjecture that through  such feedback and
mutual reinforcement, a large avalanche starting on one sandpile would have
a very high probability of inducing a (simultaneous) avalanche of
comparable magnitude on the other sandpile, despite the weakness of the
coupling. Indeed, from this scenario one might also infer  that perhaps
these avalanches would not merely be comparable in size, but in fact nearly
{\it equal} in size (that this should be the case is certainly plausible,
though, admittedly, not compellingly obvious).

While this intuitive argument is helpful, understanding the actual process
by which inter-sandpile correlations develop proves to be surprisingly
subtle and interesting. The purpose of this paper is to examine the nature
of this complex interplay between interacting automata, and to show how it
produces the observed large-scale synchronous behavior. To do so, we use a
modification of a renormalization-group procedure originally developed by
\cite{Pietronero94,Vespignani95,Hasty98} for single-sandpile models, along
with an algebraic technique.  The renormalization procedure in fact turns
out to be interesting in its own right, since it is implemented using a
Monte Carlo method which proves to be highly efficient, thus rendering
previously intractable renormalization calculations easily computable. As a
result, the methodology we employ here is likely to be applicable to a
variety of related automata problems. This paper is organized as follows.
In Section~\ref{modelsection} we present a prototype interacting-sandpile
model and describe numerical simulations which demonstrate the emergence of
large-scale synchrony. Section~\ref{RG} contains a detailed discussion of
the renormalization procedure itself and its predictions. We also describe
an alternative algebraic approach which proves useful for understanding
certain key features of the model, including the appearance of so-called
``coupling symmetry.'' Generalizations of the basic model are also
described.

\section{Basic Model and Phenomenology}
\label{modelsection}
To begin, we recall a classic sandpile model  studied by Dhar and
Ramaswamy \cite{Dhar89}.  The system consists of a two-dimensional square
lattice, where to each lattice site ij one ascribes a non-negative integer
$h(i,j)$.  The function $h(i,j)$ is called the ``height'' and represents
the number of ``sandgrains'' on a given site.  The system evolves as
follows: A lattice site is selected at random, and one grain is added to
that site. Provided the new height does not exceed a certain critical value
(taken througout this paper to be 4), then nothing further happens.  If, however, the
critical height is exceeded, then that site will ``topple''
$(h(i,j)\rightarrow h(i,j)-2)$ and spill one grain to its neighbor on the
right $(h(i,j+1)\rightarrow h(i,j+1)+1)$ and one to its neighbor above
$(h(i-1,j)\rightarrow h(i-1,j)+1)$. The affected sites on the right and
above may  in turn topple (if their heights are above the critical
threshold), and so on.  In this manner, it is possible for an avalanche
to spread across the lattice. This type of model is referred to as a
``directed'' sandpile, since an avalanche can only propagate upwards or to
the right.  Once an avalanche has exhausted itself, a new grain is dropped
onto a randomly selected site (from either sandpile), and the process
repeats.  (We mention here that the asymptotic behavior of this model remains
unchanged under a wide choice of
dropping rules, as analyzed in \cite{Dhar89}.)
 The dynamics of
this model (which is representative of a large class of related automaton
models) is surprisingly complex and has been well-documented
\cite{Dhar89,Hasty98}. One of its key features is that it exhibits
avalanches of all sizes, where ``size'' refers to the total number of
lattice sites that topple upon the addition of a single grain to the
system.

Now consider a system of two (independent) directed sandpile models, each
evolving according to the rules outlined above. The two sandpiles are
assumed to be of identical dimension, so that for every lattice site on the
first, one can associate a corresponding lattice site on the second.
(Visually, we imagine two planar lattices, one atop the other. Each site on
the top sheet is matched with the site immediately below.) We'll refer to
these two lattices as ``Sheet A'' and ``Sheet B,'' respectively. We now
allow the two sheets to interact according to the following rule: If a site
on a given sheet topples, it will, as before, always spill two grains onto
its own sheet (one to each of its  neighboring sites on the right and
above). Now, however, we assume this toppling site also has some nonzero
probability $\rho$ of spilling an additional two grains onto the other
sheet (one to each of the neighboring sites on the right and above).  In
this latter case  we require $h(i,j;s)\rightarrow h(i,j;s)-4$, where $s$
denotes the sheet on which the toppling site lies, so that sand is ``conserved.''
(Since this model is directed, it
does not matter if the updating rules for the lattice are implemented
sequentially or in parallel.) 
Note that when $\rho =0$, the two sheets are dynamically independent. Our
study  focuses  on the weak-coupling regime $(\rho << 1)$. We'll henceforth
refer to this particular model as the ``two-sheet model'' (generalizations
will be described later).

In a series of numerical simulations on this coupled system, we added
grains (one at a time) to randomly selected sites (on either sheet), and monitored the
resulting avalanches.  (Most simulations were carried out for $L=1000$ and
very few avalanches in the simulations reached the edge, so we expect edge
effects to be minimal.) For each avalanche, we tracked the number of sites
that toppled in each of the two sheets $(N_A, N_B)$, explicitly counting
multiplicities if a given site underwent multiple topples. 
A representative graph
is shown in Figure 2. The high density of datapoints along the x- and y-
axes of Fig.\ 2 for small avalanche sizes indicates that these small
avalanches remain largely confined to the sheet on which they started; only
rarely will they spill over to the other sheet. This is not  surprising,
since the two sheets are only very weakly coupled to one another
$(\rho=0.05)$ and thus the dynamics on each can be expected to be
essentially independent. However, for large avalanches, a new trend is
clearly seen to emerge: Even though, at each individual lattice site,  the
probability of a grain spilling over to the other sheet remains very low,
nonetheless a large avalanche starting on one sheet is seen to have an {\it
equally} large effect on the other sheet.  In particular, the total number
of sites that topple on each sheet (in a given avalanche) become nearly
equal in magnitude (Fig.\ 2), with a root-mean-square fractional deviation
that approaches zero (Fig.\ 3, solid line). Qualitatively, it is as though
the effective coupling strength between the two sheets increases with
spatial scale. (We will return to this point later.) We note here (for
emphasis) that had we included in Fig.\ 2 only those avalanches that
were touched off by the addition of grains to, say, Sheet A only, then
the prominent trend towards the diagonal seen in the figure for large
avalanche size would be unaffected.

Our goal is to show how this type of ``large-scale synchrony'' ({\it LSS}
for short) arises between two such weakly interacting systems. We mention
here that {\it LSS} also appears in a larger class of models than just the
numerical example described above. For instance, one can construct
``generalized two-sheet models,'' in which sites are allowed to spill
either one or two grains onto neighboring sites on either/both sheets
according to some probability matrix. As we'll show later, these
generalized systems (subject to some mild restrictions, namely, an overall
right/above symmetry) fall into the same universality class as our original
two-sheet model and hence also exhibit {\it LSS}. (We have in fact found 
that power-law behavior -- which is a  characteristic of all the sandpile
models to be discussed in this paper -- is indeed not essential for the
appearance of {\it LSS});

\section{Renormalization-Group Analysis}
\label{RG}

The fundamental behavior of these weakly interacting automata can be understood
using a renormalization-group
analysis, as we now describe. We base our work on the renormalization
 procedure developed by Hasty and Wiesenfeld \cite{Hasty98}
for (single sheet) directed-sandpile models, which
extended key work by Pietronero et al.\ \cite{Pietronero94,Vespignani95}. Adapting
this procedure for systems of interacting automata,
we show how it can be used to explain the emergence of {\it LSS}.

The basic idea behind the renormalization method is to repeatedly coarse grain the
automaton lattice into cells of successively larger sizes. In our model,
this means grouping the lattice sites on each  sheet into $2 \times 2$
blocks, then $4 \times 4$ blocks, then $8 \times 8$, and so on.  At each
stage, dynamical evolution rules are constructed which describe the
behavior of the  cells. Each time the basic cell  size is
increased, new dynamical evolution rules are constructed. The so-called
``RG map'' is a mapping that links the evolution rules for these different
cell sizes.  The behavior of the original automaton model on large spatial
scales can then  be deduced by examining the limiting behavior (i.e., fixed
points) of this RG map.
Before proceeding, we remark here upon an
important distinction between the course-graining procedure used here
for our two-automata model and the procedure used  for single-sheet models,
as described by
\cite{Hasty98,Pietronero94}. Specifically,
because we are interested in how each
sheet behaves individually, we do the coarse graining on each sheet
individually, rather than following the standard procedure which would
naturally treat the full two-sheet lattice as a single entity and coarse grain it
into cells which span both sheets.  (If this latter procedure is followed,
one finds that under the RG mapping, the two-sheet model  converges
to the single-sheet model of \cite{Hasty98,Dhar89}, proving that the two models have
the same critical exponent \cite{PaV00note}. Unfortunately, all information regarding
correlations between the two sheets is lost.)

In our model, the procedure works as follows.
Imagine that the sites on Sheets A and B have already been coarse grained
$n$ times, so that the individual
 sheets are  divided up into large ``cells,'' each comprised of
$2^n \times 2^n$ individual lattice sites.  We will use the term ``cell pair'' to
collectively refer to a cell together with its associated cell on the other sheet.
Adapting the renormalization scheme
of \cite{Pietronero94,Vespignani95,Hasty98} to our system, the evolution
rules for a cell pair can be expressed in terms of a
$3 \times 3$ probability matrix $P^n$. In particular, if a
grain is added to a cell on a particular sheet, then the associated probability
matrix is
$P^n = P^n_{\alpha,\beta}$ ($\alpha, \beta \in \{0,1,2\}$),
where $\alpha$ is the
number of grains spilled onto the same sheet (where the grain was added)
and $\beta$ is the number of grains spilled onto the other sheet. If
$\alpha=1$ (or $\beta=1$)  then the direction of the spilled grain (up or
right) is chosen at random.  If $\alpha=2$ (or $\beta=2$) then one grain is
spilled up and the other to the right. For example, in our original
two-sheet model, only spills of the types $P^0_{0,0}$, $P^0_{2,0}$ and
$P^0_{2,2}$ can occur. (This characterization of a cell's dynamics is not
quite complete. It is also  necessary to distinguish between two subcases
of $P^{n}_{1,1}$, namely, the case when the spills on Sheets A and B are in
the same direction (i.e., both to the right or both up), and when they  are
not (i.e., one to the right and one up). We'll denote these symmetric and
anti-symmetric subcases as $P^n_{1,1,s}$, $P^n_{1,1,a}$, respectively
($P^n_{1,1,s} + P^n_{1,1,a} = P^n_{1,1}$).)

The next step is to course grain the cells again
(into $2^{n+1} \times 2^{n+1}$ blocks), and construct the corresponding evolution rules governing
the new, enlarged cell pairs. In other words, we wish to determine the RG map
that relates $P^n$ to $P^{n+1}$. To do so, we utilize the procedure
developed in \cite{Hasty98} for the case of a
single-sheet automaton model. This method involves considering two-by-two
blocks of the smaller ($2^{n} \times 2^{n}$) cells, and going through all possible combinatoric
possibilities to derive the probabilities for the enlarged cells. We refer
the reader to \cite{Hasty98} for a  description of this basic
method.  There is, however, one critical
departure that we make from the procedure cited in \cite{Hasty98}. Namely,
construction of the RG map for the single automaton case in \cite{Hasty98}
proved arduous but analytically tractable (the RG map contained on the
order of 100 terms).  However, for our case of two weakly interacting
automaton, the resulting RG map is much more complex (it contains several
orders of magnitude more terms!), rendering its explicit calculation
infeasible. As described below, to surmount this difficulty we use
Monte Carlo simulations to
numerically sample the various combinatoric possibilities associated with
$P^{n}$, and in this manner can approximate the probability matrix of the
enlarged cells $P^{n+1}$. We then repeat this procedure and look at the
limiting behavior of the resulting sequence of probability matrices.

Specifically, the renormalization mapping is computed as follows.
Assume that the evolution matrix $P^n$,
which describes the cell-pair dynamics for cells of
size $2^n\times 2^n$, is known.
We now consider enlarged cells of size $2^{n+1} \times 2^{n+1}$, formed by
grouping together four  ($2^n \times 2^n$) cells  into
$2 \times 2$ blocks. The rules governing the behavior of these enlarged cells
are obtained in the following manner. Imagine dropping a single grain onto
 the lower-left subcell of a ($2 \times 2$) cell.
For sake of argument assume this cell lies on the top sheet (Sheet A).
The subcell will respond according  to rules defined by the evolution matrix
$P^n$.  For example, the
probability of that subcell not toppling is given by $P^n_{0,0}$, while the
probability of that subcell toppling onto all four of its downstream neighbors (two
on each sheet) is given by $P^n_{2,2}$. We continue to follow the avalanching process
until all subcells (in both the cell on Sheet A and the corresponding cell on Sheet B)
 are quiescent.
We now check where grains have exited the
large cell.  For example, if no grains left the large cell on Sheet A, even if there
were some internal spills, then we have an event of type $(0,0)$, while if
two grains exited onto sheet B, one in each direction and none on sheet A, then
we have an event of type $(0,2)$. However, if two grains exited on sheet B,
both in the same direction (say, to the right), and none on sheet A, then we have
an event of type $(0,1)$.  Thus, we only count the number of directions
in which grains exit the large cell, not the total number grains in each direction,
for this part of the analysis. We then repeat this procedure a large
number, of times. (Typically about $10^6$ trials are necessary  for
adequate accuracy.)  At the end of this procedure we have a unnormalized
matrix of evolution numbers, $\Lambda_{\alpha,\beta}$ which is the total number
of type $(\alpha,\beta)$ events which occurred.
However, as discussed in \cite{Pietronero94,Hasty98},
we need to ``normalize'' these probabilities in a specific manner.
In order to do this we compute $\Lambda_{0,0}$
differently than the other elements of the matrix $\Lambda$.
The procedure we use is that for each sample we take
(i.e., for each drop onto the initial subcell), we count the number of
grains that exit the large cell.  If this number is larger
than zero we add one less than this number to
$\Lambda_{0,0}$.  This ``normalization'' is very similar to the
one used by Hasty and Wiesenfeld but slightly easier to
compute in simulations (but would be more difficult analytically),
and also easier to generalize for more complex sandpiles (such as
ones in higher dimensions).  In fact, when applied to the
single sheet model studied by Hasty and Wiesenfeld, it leads to slightly more
accurate estimates of the critical exponent than their
procedure does.
(The difference between the two procedures arises when
an upper-right subcell spills two
grains in the same direction out of the large cell.
In this case our procedure adds one more to $\Lambda_{0,0}$
than Hasty and Wiesenfeld's would.)

Given the matrix $\Lambda$ it is straightforward to compute $P^{n+1}$.
 Let $|\Lambda|$ be the sum of all the elements of
$\Lambda$. We view the elements as probability amplitudes and thus we need to convert them into true probabilities to continue the renormalization procedure, thus, $P^n_{\alpha,\beta}=\Lambda_{\alpha,\beta}/|\Lambda|$.

Representative results for the renormalization process are as follows
(accurate to about $\pm 0.002$):
$$P^0=\left(\matrix{
        .500    & .000  & .000\cr
        .000    & .000  & .000\cr
        .475    & .000  & .025\cr
}\right)~
P^4=\left(\matrix{
.583    &  .067 &  .011\cr
.088    &  .105 &  .034\cr
.014    &  .038 &  .055\cr
}\right)$$
$$P^{16}=\left(\matrix{
.700    &  .001 &  .000\cr
.001    &  .185 &  .001\cr
.000    &  .001 &  .112\cr
}\right)~
P^\infty=\left(\matrix{
.702    &  .000 &  .000\cr
.000    &  .185 &  .000\cr
.000    &  .000 &  .113\cr
}\right)$$

This has three immediate consequences:

(a)  Having set the  initial probability matrix $P^0$ to correspond with
our original two-sheet model with weak coupling
($\rho = 0.05$), we find that the
RG map quickly converges to the limiting probability matrix,
$P^{\infty}$.
The principal result here is that this limiting matrix is diagonal.
This shows that in any large
avalanche (i.e., on large spatial scales)
the number of topplings on the top sheet and bottom sheet will be approximately
equal, thereby establishing the emergence of {\it LSS} in
this model.

(b) If we instead vary the values of the starting probability matrix $P^{0}$
(corresponding to the generalized two-sheet models), we find that all (nontrivial)
choices of starting configurations $P^{0}$
display the {\it same} limiting behavior $P^{\infty}$
in the renormalization analysis. Hence these generalized models
fall into the same universality
class as the original, and therefore
will all exhibit the same behavior ({\it LSS}) on large spatial scales.
In particular, this
shows, for example, that  our original two-sheet model with weak coupling
($\rho = 0.05 << 1$)  is in the same
universality class as a two-sheet model with full coupling ($\rho =1$).
In other words, when viewed on  larger and larger spatial scales,
 the weakly interacting automata begin to act as though they were very strongly
coupled. This strengthening of effective coupling constant with length scale can be
regarded as the source of the high level of correlation between the two systems.

c) If we examine the intermediate stages of the transition process $P^{0}
\rightarrow P^1 \rightarrow \ldots P^{\infty}$,
an interesting new feature appears:  Under the RG map a general starting
matrix $P^{0}$ will first become approximately symmetric (e.g., $P^{4}$)
prior to becoming nearly diagonal (e.g., $P^{16}$). (In the symmetric
phase, $P^n_{\alpha \beta} \approx P^n_{\beta \alpha}$, and ${P^n}_{11a}
\approx 0$). Hence, the renormalization analysis leads to a prediction
that, in our original two-sheet automaton model, adding a grain to a site
on, say, Sheet A, has an equal likelihood of inducing an
(intermediate-size) avalanche on Sheet B as  on Sheet A, even though the
local dynamics dictate that small avalanches are much more likely to occur
on the sheet to which the additional grain was added than on the other
sheet. This is surprising in that we started with a model in which the
coupling between neighboring lattice sites was highly asymmetric (in the
sense that each site is strongly coupled with its neighbors on its same
sheet but only weakly coupled with its neighbors on the other sheet (i.e.,
$\rho = 0.05$)),         and yet we are led to the conclusion that on
larger length scales the effective inter-sheet coupling becomes equal in
strength to the intra-sheet coupling.  A type of large scale `coupling
symmetry' has thus emerged. This prediction was tested and borne out by
numerical simulations of the automaton, as illustrated (by the dashed line)
in Fig.\ 3.

We can gain further insight into the nature of this statistical synchrony
and, in particular, the onset of this coupling symmetry, by forgoing the
above renormalization approach and instead utilizing an algebraic argument
based on work  by Dhar \cite{Dhar90} for an analogous model (see also Zhang
\cite{Zhang89}). We will take our original two-sheet model and calculate
the two-point correlation function $C(x,y)$, which describes the expected
number of topplings at site $y$, due to the avalanche caused by adding a single grain to lattice site $x$.  What we will prove is
that if two sites $x$ and $y$ are sufficiently far apart, then a symmetry
in the correlation function $C(x,y)
\approx C(x,\overline{y})$ develops, where $\overline{y}$ denotes the site
corresponding to $y$ but on the opposite sheet. 
This calculation will show that adding a grain to a given site on one sheet
will induce the {\it same} expected number of topplings on some distant 
site on its own sheet as it will  
on the corresponding (distant) site on the other sheet --
despite the weakness of the coupling between the two sheets.
(In what follows, it will be convenient to let  $x_L,x_B$ denote the
the neighboring sites immediately to the left or below a site $x$, on the same sheet.)

First, define a toppling matrix $-\Delta(x,y)$, which specifies the average
number of grains that will spill directly from a site $x$ to site $y$ in
the event that $x$ topples. (Note that here we count only {\it direct}
spillage between the two sites, not grains that might spill from $x$ to $y$
by way of intermediate sites.) For our original two-sheet model, we have $
-\Delta(y,y)=-2(1+\rho); -\Delta(y_L,y)=1; -\Delta(y_B,y)=1;
-\Delta(\overline{y_L},y)=\rho; -\Delta(\overline{y_B},y)=\rho.$ All other
components of $\Delta(x,y)$ are zero. As in Dhar \cite{Dhar90}, it is
straightforward to show that the toppling matrix and correlation function
obey the following general relation: $\sum_{z} C(x,z)
\Delta(z,y)=\delta_{x,y}$. For our model, the only terms in the toppling
matrix which contribute to the summation are the four neighboring sites of
$y$. Thus, the relationship reduces to $ C(x,y_L) + C(x, y_B) + \rho
C(x,\overline{y_L}) +
\rho C(x,\overline{y_B}) = 2(1+\rho) C(x, y).
$
We observe, however, that this
relation is precisely the formula for the
probability that a certain random walk starting at site $x$ will hit site $y$.
In this random walk, at every step the walker is equally likely to go up or right,
and switches between sheets with probability $\rho/(1+\rho)$.  It follows then that
the probability that a walker starting on one sheet will end up  on
that same sheet
$k$ steps later is $(1+[(1-\rho)/(1+\rho)]^k)/2$.
Since this approaches
$1/2$ for large $k$, we conclude
that $C(x;y) \approx C(x;\overline{y})$ for $x$ and $y$
sufficiently far apart.
(More precisely, the fractional difference between
these correlations scales like
$[(1-\rho)/(1+\rho)]^{k}$, where $k$ is the distance between sites $x$ and $y$
in the ``taxicab'' metric, assuming of course that $y$ is reachable from $x$, else both correlations are $0$.)
Hence, this demonstrates that on sufficiently large spatial scales,
the intra-sheet and inter-sheet coupling
become equal.

\section{Conclusions}
In summary, we have examined (in the context of a few specific examples)
the nature of the complex correlations arising between weakly interacting
automata, and have used a Monte-Carlo implementation of a renormalization
group analysis to understand the appearance of large-scale
statistical synchrony in these systems. Since both our methods of analysis
and the properties of SOC systems are extremely robust (e.g., the extension
of the algebraic analysis to more general models is straightforward), we
believe that  the types of inter-sandpile correlations found here will
likely be a generic feature of  other weakly coupled  SOC  systems.
In fact,
preliminary analysis suggests that these properties even arise  in some automata
models which do not exhibit SOC, such  as dissipative  models. We note that
our Monte Carlo approach for studying the RG map turns out to
be remarkably efficient and may in fact provide the key to applying
renormalization to more complex automaton models.

Lastly, we remark that the emergence of strong statistical correlations described here
 raises a number of interesting questions, including
(i) Is there some universal scaling law describing how the
length scale at which strong correlations arise varies with the inter-sheet
coupling strength; and (ii) Might it be possible to recast
this phenomenon as a type of phase
transition that occurs with increasing spatial scale?

We would like to thank D. Dhar for helpful comments on the manuscript.

\newpage

\section{Figure Captions}

Figure 1. A representative time series. Shown is the total number of topples on each sandpile for each in a series of avalanches. Note that large peaks occur simultaneously and are approximately equal in
magnitude, while smaller peaks are relatively uncorrelated in both time and size.
(The dataset was
generated from an automata model described in Section~\ref{modelsection}.  Note that for
illustrative purposes, we have added
one to the avalanche sizes in order to avoid singularities associated with the logarithmic scaling in the plots.)

Figure 2.  Large-scale synchrony.  The number of topples on each sheet
during avalanches
in the two-sheet model with
coupling parameter $\rho=0.05$ is shown.
Observe that for large avalanche sizes strong correlations
develop, with $N_A$ and $N_B$ becoming approximately equal.

Figure 3.  The  root-mean-square fractional  deviation
$f_{rms}$ (solid line) between $N_A$ and $N_B$
vs.\  avalanche size $(N=N_A+N_B)$ for the
data shown in Fig.\ 2. The decrease  in  $f_{rms}$ with  size indicates that, on
large length scales, the two sheets behave as though they were strongly coupled.
A related phenomenon, `coupling symmetry' (see text), is illustrated by
the dashed curve showing the average fractional deviation $(f_{ave}=\langle (N_A-N_B) /(N_A+N_B) \rangle)$,
where the average is computed over only those avalanches that were initiated by the addition of
one grain to Sheet A.  Observe that such avalanches, if small, remain primarily confined to
Sheet A (as expected), while large ones divide equally between the two sheets (since
$f_{ave} \rightarrow 0$).

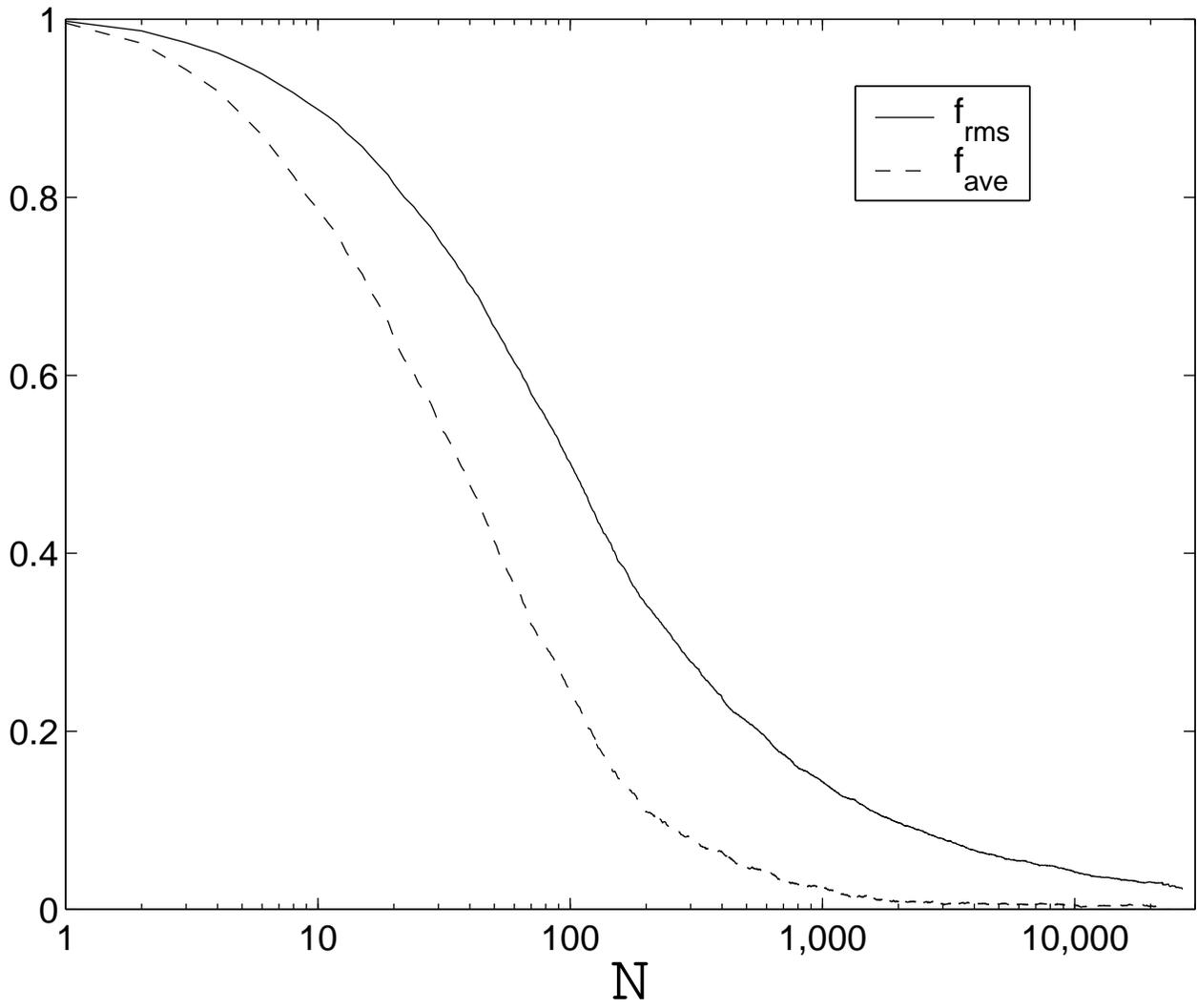
\end{document}